# Adaptive, Closed Loop OFDM-Based Resource Allocation Method using Machine Learning and Genetic Algorithm

Wafaa S. Taie[1], Ashraf H. Badawi[2], Ahmed F. Shalash[1]
[1]Electronics and Communications Engineering, Cairo University, Giza, Egypt
[2]Zewail City of Science and Technology, Giza, Egypt.
{wafa.taie@gmail.com, abadawi@zewailcity.edu.eg, shalash@ieee.org}

*Abstract*— in this paper, the concept of Machine Learning (ML) is introduced to the Orthogonal Frequency Division Multiple Access-based (OFDMA-based) scheduler. Similar to the impact of the Channel Quality Indicator (CQI) on the scheduler in the Long Term Evolution (LTE), ML is utilized to provide the scheduler with pertinent information about the User Equipment (UE) traffic patterns, demands, Quality of Service (QoS) requirements, instantaneous user throughput and other network conditions. An adaptive ML-based framework is proposed in order to optimize the LTE scheduler operation. The proposed technique targets multiple objective scheduling strategies. The weights of the different objectives are adjusted to optimize the resources allocation per transmission based on the UEs demand pattern. In addition, it overcomes the trade-off problem of the traditional scheduling methods. The technique can be used as a generic framework with any scheduling strategy. In this paper, Genetic Algorithm-based (GA-based) multi-objective scheduler is considered to illustrate the efficiency of the proposed adaptive scheduling solution. Results show that using the combination of clustering and classification algorithms along with the GA optimizes the GA scheduler functionality and makes use of the ML process to form a closed loop scheduling mechanism.

I. INTRODUCTION

Multi-user diversity is one of the most important features that OFDMA provides to Worldwide Interoperability for Microwave Access (WiMAX) [1] and LTE systems [2]. In this work, we consider an LTE system and focus on Radio Resources Manager (RRM) which is responsible for UEs scheduling function in the enhanced NodeB (eNB). RRM manages to accept or reject the UE bandwidth request through Admission Control (AC) and schedules the UEs that are previously admitted by AC using Packet Scheduling (PS). In The LTE downlink (DL) transmission, the frame structure is based on OFDMA technology, where a UE can be assigned one or more resource block (RB) according to its traffic demand and the scheduler strategy.

In addition to the multiuser diversity gain, the CQI provided by the LTE physical layer highly impacted the radio resources allocation strategies, where some of the conventional LTE schedulers managed to achieve the highest possible data rates for UEs and increased the overall eNB capacity [3]. However, the CQI is a single value, lumping all information the UE is reporting about the quality of the channel. The UE uses a combination of signal-to-noise ratio, signal-to-interference plus noise ratio, and signal-to-noise plus distortion ratio to estimate CQI. Closing the loop with the eNB with UE-reported instantaneous user throughput can enhance the resource allocation.

Additionally, the LTE scheduling system may target other objectives to optimize besides the throughput such as "fairness" [4] [5], "latency" [6]…etc. Therefore, in order to compromise and overcome the tradeoffs and achieve optimum resource allocation, the scheduling function tends to be an optimization problem with multiple objectives. In addition to the increasing number of mobile devices and real-time services, more challenges are added to the scheduler operation in high mobility environment to meet UEs demands and QoS as well as Quality of Experience (QoE). Moreover, the CQI, which is reported through the uplink (UL) transmission to be utilized in the DL, does not efficiently reflect the Mobile UE (MUE) fast varying channel conditions.

Having the scheduler unaware of the MUE channel conditions, in addition to the strict QoS requirements of real-time services, the conventional scheduling methods cost high computational complexity to achieve the optimum allocation patterns. Therefore, heuristic techniques (such as GA) have been proposed to solve scheduling problems that are mathematically complicated and require high computational power. GA has a large search space and can converge to a suboptimal solution with acceptable computational complexity compared to mathematical solutions, such as linear programming. The systems basically "learns" the effects of the different allocation schemes via closing the loop at the application layer and thus enhances the scheduling process.

Genetic Algorithm, first introduced by John Holland [7], is a rapidly growing area of the evolutionary computing field of artificial intelligence. It is a search that determines the optimum solution set representing the newer generations. The search begins from a randomly generated population and evolves over successive iterations. GA applies three main functions to propagate its population from a generation to the other. The selection function chooses the fittest individauls according to the fitness function objective to be the parents of the next generation. It tries to ensure quality of the individuals while the algorithm progresses. The crossover then propagates features of good surviving individuals from the current population into the future population. Finally, in order to promote diversity in population characteristics, mutation is applied. It also allows for global search of the population space and prevents the algorithm from getting into local minima.

Recently GA has been widely adopted in wireless networks resources allocation. Authors of [8] asserted that applying GA to resource allocation in OFDMA-based systems is efficient and outperforms the traditional scheduling techniques in satisfying UEs QoS requirements. Moreover, [9] explains how GA scheduler approaches the optimal resource allocation and surpasses traditional allocation methods. GA is also used to enhance cell edge users as well as the overall network performance [10]. GA-based quality enhancement approach significantly enhances video quality and satisfies the delay bound as explained in [11]. In [12], the GA scheduler can be effectively used to manage throughput and fairness objectives in dynamic network scenarios. However, the main disadvantage of GA-based schedulers is the computational complexity compared to the legacy scheduling schemes.

In this paper, LTE scheduler utilizes ML techniques to address the previously highlighted scheduling challenges. A clustering technique is used to process the big data that the eNB collects from the UEs attached to it. This data, when used for training the scheduler on how a specific allocation pattern worked in previous cycles, given the channel conditions and demand pattern from UEs at that time. In addition, a classification technique teaches the scheduler how to deal with the different users' traffic demands. Classifying the UE demand to the correct group helps tune the scheduler different objectives and optimize its performance. In addition, the proposed technique introduces a solution that significantly decreases GA scheduler computational complexity.

The paper is organized as follows. Section II describes the proposed algorithm model. Section III provides the simulation environment and parameters and the evaluation of the proposed technique's performance.

## II. Adaptive Resource allocation Technique in OFDM-Based systems Methodology

### A. Problem Statement

Due to the high mobility supported in LTE and LTE advanced, the scheduling function has to be intelligent enough and capable of meeting the real time QoS requirements. These factors are reflected on the scheduler computational complexity. GA is used as efficient solver to complex objective functions with low computational cost. In our proposed methodology, GA is used as an alternative to linear programming for solving the linearized model of the Maximum Throughput (Max. TP) scheduler proposed in [13]. In our implementation, encoding the GA chromosome as vector of integers managed to model the constraints without the need to the uni-modular matrix A, and therefore the objective function mathematically is given in (1).

$$f1 = \max \sum_{m=1}^{M} \sum_{n=1}^{N} C(m,n)\, a(m,n) \quad (1)$$

Constrained to:
- $a(m,n) = \begin{cases} 1, \text{if } n \text{ is assigned to user } m \\ 0, otherwise \end{cases}$

Where each row in the matrix C represents the efficiency of a user over all the N resource blocks according to the provided average CQI value presented in [13].

Our suggested adaptive scheduler model is a weighted multi-objective function given in (2)

$$Max\ [w1f1 - w2f2] \quad (2)$$

Where $w1$ and $w2$ are the adaptive weights, $w1 + w2 = 1$.

The objective function $f1$ targets rate maximization [13] described in (1), while objective function $f2$ that aims at meeting the Guaranteed Bit Rate (GBR) UEs throughput demands described in (3).

$$f2 = min \sum_{All\ GBR\ UEs}(\sum_{n=1}^{N} Rm(n)\ a(m,n) - R_q(m)) \quad (3)$$

Where $R_m(n) = \sum_{i=1}^{q(m,\max(j))} b(m,i) * r(i)$.
Constrained to:
- $a(m,n) = \begin{cases} 1, \text{if } n \text{ is assigned to user } m \\ 0, otherwise \end{cases}$
- $r(m) \geq R_q(m)$, to guarantee minimal data rate for each user $m$
- $\sum_{i=1}^{q(m,\max(j))} b(m,i) = 1$, all RBs belongs to a user $m$ have the same modulation and coding scheme (MCS) index (LTE constraint).

Where, $r(i)$ is the data rate per single RB, $i$ is the MCS index, $j(m,n)$ is the channel CQI (Channel Quality Indicator) of user $m$ over $N$ RBs, $q(m,\max(j))$ is the highest MCS can be supported by $m$ and $b(m,i) = \begin{cases} 1, \text{if MCS } i \text{ chosen by user } m \\ 0, otherwise \end{cases}$

### B. The Proposed Algorithm Description

The idea of the suggested technique is to collect the users' information and demands pattern that are sent to the eNB and storing them on a database for further processing as a big data using ML techniques. Massive amount of information about the similarities and relationships between users' data patterns can be identified and utilized for the LTE network performance optimization, and for providing the user better QoE. The proposed technique is considered as a framework and can be utilized by different scheduling strategies. The objective of the proposed technique can be summarized as follows:

*1) Providing a feedback to the scheduler objective functions about the trend of the users requests (in the form of tunable weights to adjust the scheduler strategy)*

*2) Optimizing the GA- based scheduler operation time. Since the latency is an essential factor in the scheduler operation.*

The proposed framework is described in Fig. 1. In the data collection and analysis stage, the UEs demands, which are requested from the eNB over time, are stored in the "UEs Demand Patterns database". These demand patterns are clustered periodically to (K) clusters using Kmeans clustering technique [14]. For LTE network of (M) UEs and (N) resource blocks, the UEs' Demand vector can be represented as:

$$D = [R_1, R_2 \ldots R_m \ldots, R_M],$$

Where $R_m$ is the throughput requested by the UE of index $m$ and $m = 1: M$.

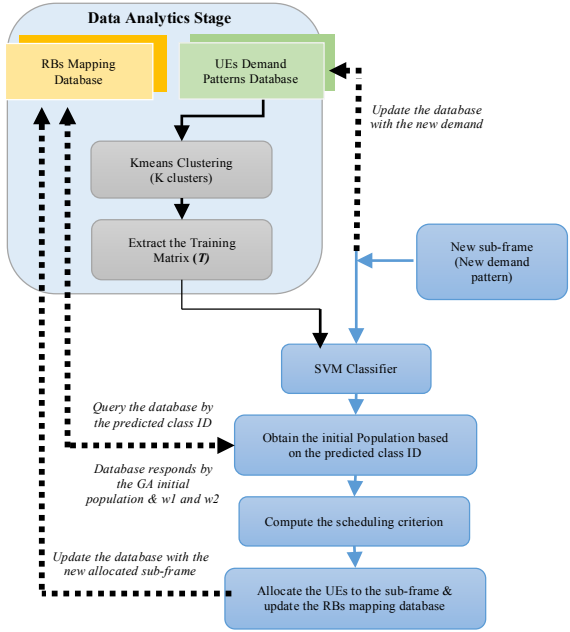

Fig. 1 the proposed adaptive scheduling algorithm block diagram

The data analytics stage is considered offline and its key function is to prepare the training matrix (T) that the scheduler will utilize in its operation. The Training Matrix (T) is illustrated in Fig. 2.

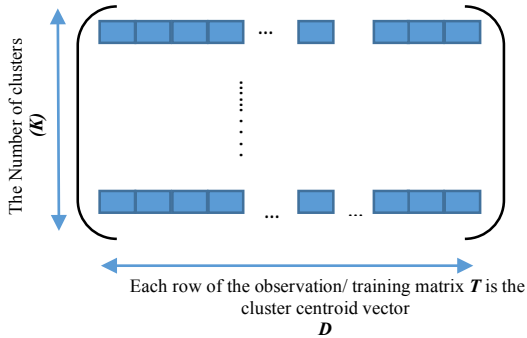

Fig. 2 the SVM training matrix description

The proposed adaptive scheduler operation is per sub-frame (i.e. when a new UEs demand request arrives to the eNB and already admitted). The scheduler classifies the new demand pattern, using Support Vector Machine (SVM) classifier [15] [16], to the closest cluster of demands. According to the cluster index, the adaptive weights w1 and w2 are adjusted.

In the case when the scheduler is GA-based, as in our implementation, the proposed framework can optimize the GA time complexity. Due to the similarities in users' requirements over time; eNB sees some similar UEs' demands over a period of time and its scheduler allocates UEs to the frame resources successfully, hence eliminating the need to perform the same repeatedly for demand patterns (i.e. in some cases the GA gets to the optimum allocation pattern in the first generation if it was fed the appropriate initial population). Therefore, if eNB manages to exploit the UEs requests similarities (using clustering) and classifies the new traffic request to a suitable cluster, it would significantly decrease GA complexity and pass to it an initial population that is close to the optimum allocation pattern for the LTE frame. The RBs-UEs mapping patterns database function is to store the optimum allocation patterns for the previous sub-frames and their corresponding fitness values. The RBs-UEs mapping patterns database has number of entries equals to the number of the UEs demand patterns clusters (K).

Each entry can be represented as a vector $Z = [z_1, z_2, ... z_n ..., z_N]$, where $z_n$ represents the scheduled UE index and $n$ is RB number, $n = 1,2,... N$.

Depending on the group that the new demand is classified to, the corresponding RBs-UEs mapping pattern can be fed to the GA as initial population. Results illustrate how the initial population impact the GA performance.

After the scheduler accomplishes its operation, the RBs-UEs mapping pattern database is updated with the optimized new pattern. It's obvious that the performance of the GA time complexity can be optimized as the number of sub-frames increases.

In order to implement the suggested scheduling strategy given in (3) using the proposed ML-based solution, three cases are considered in the weights adaptation mechanism as described below:

- All UEs are GBR UEs, in this case $w1=0$ & $w2=1$
- All UEs are non-GBR UEs, where $w1=1$ & $w2=0$
- Mixed case; where some UEs are GBR UEs and some are non-GBR, in this case $w1 = 1 - w2$ & $w2 = \frac{Number\ of\ GBR\ UEs}{M}$.

III. THE PERFORMANCE EVALUATION

A. System Model and Simulation Parameters

The LTE System Level Simulator *v1.6r885* by Institute Of Communication and Radio Frequency Engineering, Vienna University of Technology, Vienna [17] is used. LTE system and GA parameters are shown in Table I.

TABLE I. SIMULATION PARAMETERS

| Frequency (GHz) | 2.14 |
|---|---|
| Bandwidth(MHz) | 5 |
| Number of Resource Blocks | 25 |
| Number of eNBs | 1 |
| Number of User Equipment (UE) per eNB | 25 |
| Number of Transmission Time Intervals (TTIs) | 20 |
| UE Speed (km/h) | 5 |
| Number of Transmitters | 1 |
| Number of Receivers | 1 |
| Antenna Type | Omni-directional |
| Transmission Mode | SISO |
| Transmission Time Interval (ms) | 1 |
| GA Population size | 100 |
| Number of generations limit | 200 |

B. Results

In this section, different simulation scenarios are conducted to evaluate the performance of the proposed adaptive scheduling technique. Three cases of different UEs demand patterns are investigated. Simulation results are provided to illustrate the impact of tuning $w1$ & $w2$ in each case. The performance of the proposed methodology is compared to the following schedulers: (1) Max. TP (2) Adaptive GA-based Max. TP (3) Proportional Fair (PF). Jain's fairness index [18] and the UEs average satisfaction ratio are the performance evaluation metrics, in addition to the UE throughput statistics. Jain's fairness index $f(r_1, r_2, …, r_M)$, as given in (4), measures the ratio of the number of users that are equally share the resources to the total number of users.

$$f(r_1, r_2, …, r_M) = \frac{(\sum_{i=1}^{M} r_i)^2}{M \sum_{i=1}^{M} r_i^2} \quad (4)$$

Where $M$ represents the number of users and $r_i$ is the throughput of user $i$. The average GBR UEs satisfaction ratio, given in (5) measures the ability of the scheduler to meet the GBR UEs demands.

$$g(r_1, r_2, …, r_M) = \frac{1}{M} \sum_{i=1}^{M} \left(\frac{r_i}{R_i}\right) \quad (5)$$

Where $M$ represents the number of GBR UEs, $r_i$ is the throughput of user $i$ and $R_i$ is the throughput demand of user $i$. Fig. 3 depicts the case of a non–GBR UEs demand pattern when the weights are directly adjusted to $w1$=1 & $w2$=0. PF and Max. TP technique, as well-known schedulers of a similar strategy, are added to the comparison. The adaptive GA-based Max. TP scheduler performs better than Max. TP and PF regarding the Peak UE throughput. It also provides higher cell-edge throughput values compared to Max.TP due to the heuristic approach of GA-based scheduler, however it provides quite less average UE throughput. Fig. 4 depicts Jain's fairness index for all of the scheduling techniques. In this case where $w2$ =0, the GA-based technique objective, like the Max. TP, is the throughput maximization, given in (2), at the expense of achieving fairness. However, it provides slightly higher fairness percentage than Max.TP.

According to the proposed adaptive scheduler strategy, if there exist GBR UEs, $w2$ is directly adjusted to

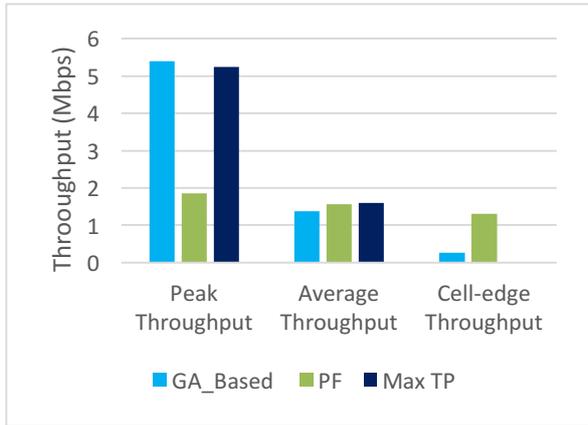

Fig. 3 UE throughput comparison between the adaptive GA-based scheduler and different scheduling strategies (bandwidth 10 MHz)

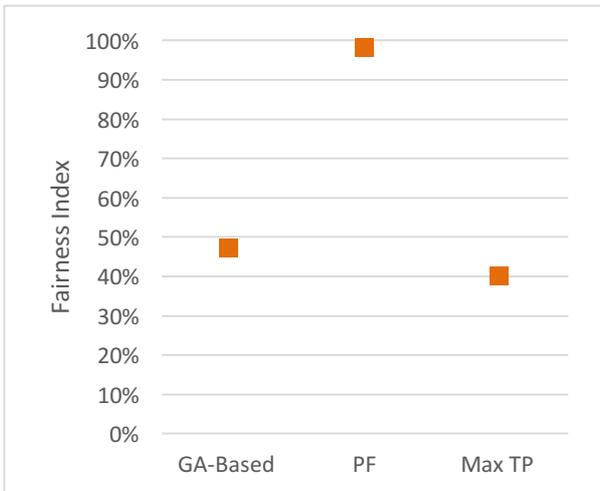

Fig. 4 the proposed GA- based scheduling fairness index compared to Max. Throughput scheduler in case of non-GBR UEs (**$w1$**=1 & **$w2$**=0) values greater than 0. Fig. 5 provides statistics for the case of mixed UEs demand pattern, 50% of the users are GBR UEs then the weights are adapted to $w1$ = 0.5 & $w2$= 0.5. As depicted, the proposed GA-based scheduler guarantees 100% average satisfaction level for all the GBR UEs compared to the Max. TP and PF. Moreover, The UE peak throughput value, shown in Fig. 6, highlights the change that happened in the scheduler strategy based on the provided feedback. The weight of the throughput maximization term $w1$ is decreased from 1 to 0.5 which resulted in less UE peak throughput value compared to Fig. 3. In Fig. 3 and Fig. 6, the different bandwidth values do not affect the performance of the proposed technique compared to other scheduling methods. However, it is highlighted to explain the 50% drop in the throughput values of Fig. 6.

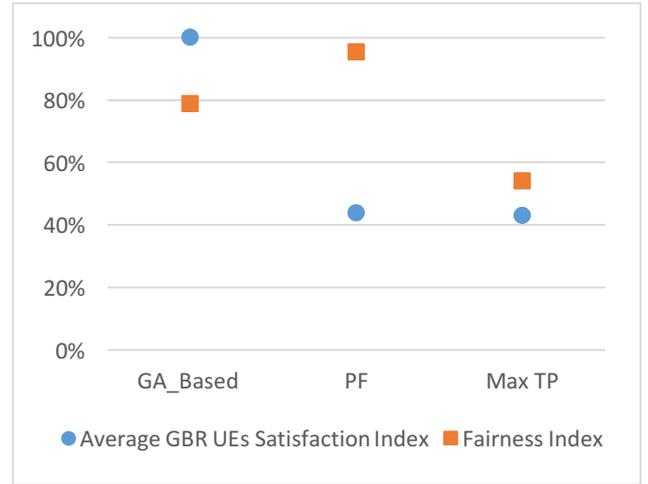

Fig. 5 GBR UEs average satisfaction index in case of mixed GBR and Non-GBR demand pattern

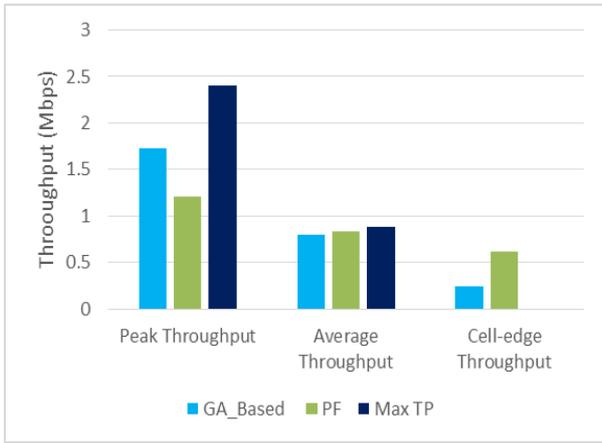

Fig. 6 UE throughput statistics in case of mixed GBR and Non-GBR demand pattern (bandwidth 5 MHz)

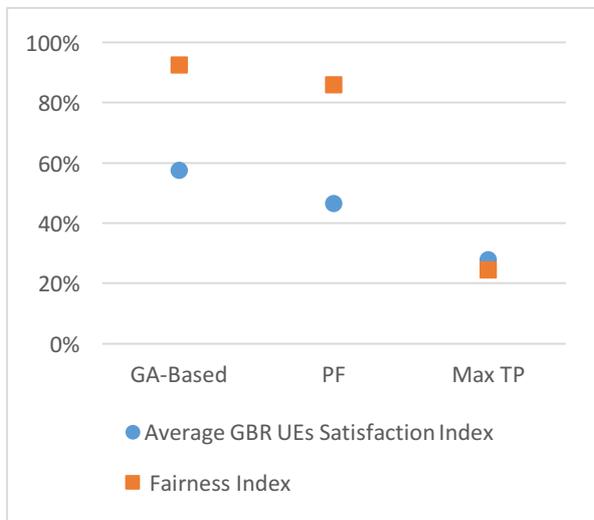

Fig. 7 average GBR UEs satisfaction index in case of high mobility and all GBR UEs demand

In order to investigate the proposed algorithm behavior in high mobility conditions, a simulation scenario is conducted with UE speed up to 200 km/h. In addition, the third type of UEs demand patterns is provided, where all UEs are GBR UEs while the objective function weights are tuned to $w1=0$ & $w2=1$. In addition to that, the number of UEs per eNB is increased to 30 UE in order to model the case of congestion and high demand. Results provided in Fig. 7 depicts that the proposed GA-based technique is more robust to the fast channel variations compared to the Max. TP technique. It managed to satisfy the GBR UEs demands with a percentage up to 60% while achieving a fairness percentage up to 92%, which significantly out performs the Max. TP. Despite the mobile channel conditions, the proposed algorithm behavior in this case is close to the PF techniques as Fig. 7 illustrates. The proposed scheduler strategy increases the UEs throughput to meet their demands while maintaining a certain fairness level during the resources assignment to the competing users.

Fig. 8 illustrates that the adaption decision can be optimized using the trends of different performance metrics with respect to the weights $w1$ and $w2$. The fairness index is chosen as a performance metric. The behavior of the fairness index depicts that when the value if $w1$ is low (close to 0) the proposed technique objective is to satisfy the different UEs' demands equally, and behaves quite similar to the PF scheduler. While $w1$ is close to 1, the proposed algorithm acts similar to Max. TP technique. And the intermediate values of $w1$ yield performance in between as given in (2), which proves the adaptation strategy of the proposed technique. Therefore, the trend of different performance metrics can be utilized as a design parameters for the scheduler feedback and objective function.

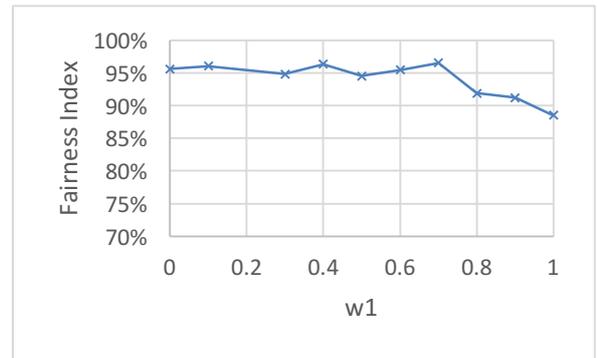

Fig. 8 Jain's fairness index behavior vs. the proposed scheduler adaptive weights

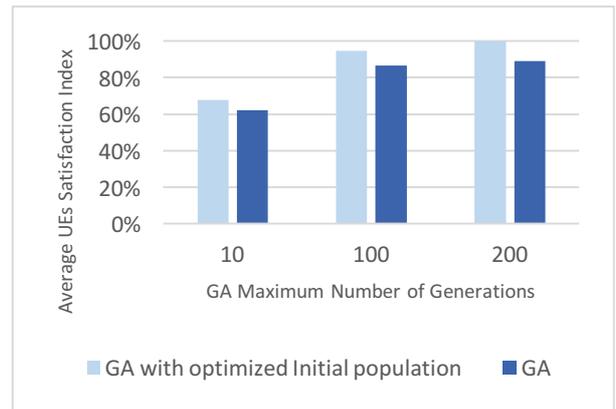

Fig. 9 comparison between The GA-based algorithm with and without optimizing its initial population

Finally, Fig. 9 shows how the proposed framework using machine learning contributes in enhancing the GA operation. The GA time complexity is the main disadvantage of the algorithm to be adopted in LTE schedulers. As the GA is population based and the number of generations, utilized in order to reach an optimal or suboptimal fitness value, consumes a major part of the algorithm operation time. As depicted, providing the GA with an optimized initial population can significantly improves the scheduler performance compared to the GA-based scheduler performance with a random initial population considering the same number of generations. Or in other words, GA with an optimized initial population can guarantee the same performance with less number of generations. In the proposed technique, the efficiency of the ML techniques, which are used in data clustering and demand pattern classification, reflects directly on the GA convergence time.

*C. Numerical Efficiency Discussion*

This section illustrates the key benefit from using the machine learning techniques in combinations with the GA scheduler, namely the improved numerical efficiency of the modified GA scheduler. The traditional GA algorithm order of numerical complexity formula is given in [8]: "$G \cdot [O(3Ln + 2L \log(2L))]$" where, $G$ is the max number of generations and $L$ is size of population pool, and $n$ is number of RBs. The optimized GA scheduler time complexity is reduced by a multiplier factor ($G$). Kmeans complexity of order $O(KNdI)$[15] and SVM classifier maximum complexity is $O(d.K^3)$[16], where $K$ is the number of clusters, $N$ represents the number of demand patterns the eNB stores, $d$ represents the demand vector length. Kmeans, the clustering algorithm, needs to run online during the training phase, while during the scheduler operation, it runs offline periodically to update clusters with the latest demand patterns. Accordingly, the proposed scheduler numerical complexity is the sum of SVM classifier complexity and the GA complexity. As $K$ and $d$ are adjustable parameters; $K$ is chosen to guarantee efficient clustering according to the scheduler strategy and according to the network capacity, the operator can control the value of $d$. While G and L are set to large values to optimize GA functionality. Hence, the additional computations required by the SVM algorithm will always be less than or equal to the full GA algorithm. Therefore, the combination of Kmeans and SVM algorithms with GA will be more numerically efficient than GA alone.

## IV. CONCLUSION

In this paper, a framework for adaptive LTE scheduling strategy using ML is proposed. ML is successfully utilized for the scheduling function optimization and providing a real-time feedback to the LTE scheduler. The UEs demand patterns are processed and analyzed using Kmeans clustering and SVM classifier and accordingly the scheduler objective function is adjusted. The simulation results demonstrated that the proposed technique can solve the traditional schedulers' tradeoff problem on a sub-frame basis. Therefore, each cluster of similar UE demand patterns is treated with a different scheduling strategy. GA is adopted in the proposed scheduler, in order to highlight that the heuristic method of GA achieves better performance than conventional schedulers in unpredictable channel conditions. The LTE GA-based schedulers' time complexity is optimized by significantly decreasing the maximum number of generations consumed to achieve a certain performance level. Simulation results proved that the proposed GA-based scheduler is remarkably efficient in case of high mobility channel conditions. In addition, more optimization can be done in clustering and classification algorithms to achieve more efficient versions of the suggested framework. For example in our model we assume that number of UEs is constant and hence the observation vector length in clustering is of constant length. Further work needs to be done to reach an optimal approach for setting the cluster size and the feedback provided to the scheduler.